\documentclass[prd,
twocolumn,
showpacs,preprintnumbers,amsmath,amssymb,
superscriptaddress,
a4paper,nofootinbib,longbibliography]{revtex4-2}

\usepackage{setspace}

\usepackage{nicefrac}

\usepackage{esint}

\usepackage{empheq}
\usepackage[most]{tcolorbox}

\usepackage{natbib}
\usepackage{amssymb,amsbsy,amsmath,amsfonts}
\usepackage{graphicx}
\usepackage{epsf,epsfig,float,latexsym,amsthm,fancyhdr,rotating}
\usepackage{graphics,psfrag,longtable}
\usepackage{slashed}
\usepackage[normalem]{ulem}
\usepackage[colorlinks,citecolor=blue,linktoc=all,linkcolor=cyan,urlcolor=blue]{hyperref}
\usepackage{upgreek}

 \usepackage{multirow}

\def\beq{\begin{equation}}
\def\eeq{\end{equation}}
\def\bea{\begin{eqnarray}}
\def\eea{\end{eqnarray}}
\def\beqa{\begin{equation}\begin{array}{l}}
\def\eeqa{\end{array}\end{equation}}
\def\eqlab#1{\label{eq:#1}}

\def\eref#1{(\ref{eq:#1})}
\def\Eqref#1{Eq.~(\ref{eq:#1})}

\def\fref#1{\ref{fig:#1}}
\def\Figref#1{Fig.~\ref{fig:#1}}

\def\barr{\left(\begin{array}{c}}
\def\earr{\end{array}\right)}
\def\bmat{\left(\begin{array}{cc}}
\def\emat{\end{array}\right)}
\def\al{\alpha}

 \def\De{\Delta}

\def\si{\sigma}

\def\nn{\nonumber}
\def\dd{\mathrm{d}}

\DeclareMathOperator\im{Im}

\def\3d{3-D}

\def\ol#1{\overline{#1}}

\def\piEFT/{$\slashed{\pi}$EFT}

\def\2PE{2$\upgamma$}

\usepackage{xcolor}
\usepackage{bm}
\usepackage{slashed}
\usepackage{graphicx}
\allowdisplaybreaks

\makeatletter
\g@addto@macro\bfseries{\boldmath}
\makeatother

\begin{document}

\author{Volodymyr Biloshytskyi}
\affiliation{Institut f\"ur Kernphysik,
 Johannes Gutenberg-Universit\"at  Mainz,  D-55128 Mainz, Germany}
 \author{Iulian Ciobotaru-Hriscu}
\affiliation{Institut f\"ur Physik,
 Johannes Gutenberg-Universit\"at  Mainz,  D-55128 Mainz, Germany}
\author{\\Franziska Hagelstein}
\affiliation{Institut f\"ur Kernphysik,
 Johannes Gutenberg-Universit\"at  Mainz,  D-55128 Mainz, Germany}
\affiliation{Paul Scherrer Institut, Forschungsstrasse 111,
5232 Villigen PSI,
Switzerland}
\author{Vadim Lensky}
\affiliation{Institut f\"ur Kernphysik,
 Johannes Gutenberg-Universit\"at  Mainz,  D-55128 Mainz, Germany}
\author{Vladimir Pascalutsa}
\affiliation{Institut f\"ur Kernphysik,
 Johannes Gutenberg-Universit\"at  Mainz,  D-55128 Mainz, Germany}

\title{The QED of Bernab\'eu-Tarrach sum rule for electric polarizability 
and its implication for the Lamb shift}

\begin{abstract}
We attempt to rehabilitate a sum rule (proposed long ago by Bernab\'eu and Tarrach) which relates the electric polarizability of a particle to
the total photoabsorption of quasi-real longitudinally polarized photons by that particle. We discuss its perturbative verification in QED,
which is largely responsible for the scepticism about its validity. The failure of the QED test can be understood via the Sugawara-Kanazawa
theorem and is due to the non-vanishing contour contribution in the pertinent dispersion relation. We show another example where
this contribution is absent and the perturbative test works exactly. On the empirical side, we show that the sum rule gives a reasonable
estimate of the $\pi N$-channel contribution to the proton electric polarizability. If this sum rule is valid indeed, there should be a
sum rule for the so-called ``subtraction function'' entering the data-driven calculations of the polarizability effects in the Lamb shift.
We have written down a possible sum rule for the subtraction function and verified it in a perturbative calculation.
\end{abstract}

\date{\today}

\maketitle

\section{Introduction}

In 1975 Bernab\'eu and Tarrach published a sum rule for the electric dipole polarizability $\al_{E1}$
of a spin-1/2 particle \cite{Bernabeu:1974mb}, 
\beq 
\eqlab{BTsr}
\al_{E1} - \frac{\al_\mathrm{em} \varkappa^2}{4M^3} = \frac{1}{2\pi^2} \int_0^\infty \dd \nu \left[ \frac{\sigma_L(\nu,Q^2)}{Q^2} \right]_{Q^2\to 0} \,,
\eeq
where the second term involves the anomalous magnetic moment of the particle and its mass, $\varkappa$ and $M$; $\al_\mathrm{em}\simeq 1/137$ is the fine-structure constant.  The right-hand-side 
is given by the energy-integrated total longitudinal-photoabsorption cross section $\sigma_L$, function of
the photon energy $\nu$ and virtuality $Q^2$, taken in the limit of $Q^2\to 0$.\footnote{The smooth $Q^2\to 0$ limit
of $\sigma_L/Q^2$ is guaranteed by electromagnetic gauge invariance.}
This sum rule is a ``virtual sibling" of the celebrated Kramers-Kronig relation written for real photons, and, more specifically, of the Baldin sum rule, for the sum of electric and magnetic polarizabilities \cite{Baldin:1960}:
\beq 
\eqlab{Bsr}
\al_{E1}  + \beta_{M1} = \frac{1}{2\pi^2} \int_0^\infty \dd \nu \frac{\sigma_T(\nu)}{\nu^2} ,
\eeq
which involves the cross section of total photoabsorption of real (transverse) photons $\sigma_T$.

The Baldin sum rule has been instrumental for the data-driven evaluations of the nucleon polarizabilities, and
since long \cite{Damashek:1969xj} provides the most stringent empirical constraint on the sum of proton polarizabilties, 
see \cite{Gryniuk:2015aa} for the state of the art.  In contrast, the Bernab\'eu-Tarrach (BT) sum rule for nucleons was discarded \cite{Lvov:1998csf} (more recently, in \cite{Gasser:2015dwa}); its use for nuclei was discussed in
\cite{Bernabeu:1998wf}. 
Llanta and Tarrach \cite{Llanta:1978xd} were first to discredit the sum rule by showing that it fails
a perturbative verification in leading-order Quantum Electrodynamics (QED). This calculation will be revisited here (Sec.~\ref{sec:deriving}) and complemented by an analogous calculation
in chiral perturbation theory ($\chi$PT).
Our main conclusion is that the BT sum rule is valid, if convergent. 
This means the proton electric and magnetic polarizabilities could be evaluated separately using the two sum rules,
which would be extremely interesting in context of the current controversy in determination
of these polarizabilities via the Compton scattering experiments at HI$\gamma$S \cite{Li:2022vnz} versus MAMI \cite{A2CollaborationatMAMI:2021vfy}. 

Another important implication of the valid BT sum rule would be
the possibility of a data-driven evaluation of the ``subtraction-function contribution'' to the proton-structure effects in the Lamb shift of
muonic hydrogen. This contribution brings one of the significant uncertainties in the extraction of the proton charge radius from muonic hydrogen spectroscopy 
\cite{Pohl:2010zza,Antognini:2012ofa,Antognini:1900ns,Carlson:2011zd,Birse:2012eb}, see also \cite{Antognini:2022xoo,Pachucki:2022tgl}
for the most recent reviews.

\section{Deriving the sum rule and QED failure}
\label{sec:deriving}

The BT sum rule  can be derived from general properties of the forward doubly-virtual Compton scattering (VVCS), as reviewed,  e.g., in 
\cite{Drechsel:2002ar, Pascalutsa:2018ced,Hagelstein:2015egb}.
We follow Ref.~\cite[Ch.~5]{Hagelstein:2015egb}. 
Considering the amplitude for the VVCS of longitudinally polarized photons, $T_L (\nu, Q^2)$, and using its analytic properties in the complex $\nu$-plane 
one can write down a dispersion relation which, by means of the optical theorem (unitarity),\footnote{Our convention for the virtual-photon flux is such that 
$\im T_L(\nu, Q^2) = \nu \sigma_L (\nu, Q^2)$, for any $Q^2$.} involves the inclusive photoabsorption cross section $\si_L$, 
\beq 
\eqlab{TLDR}
T_L (\nu, Q^2) = \frac{2}{\pi} \int_{\nu_0}^\infty \dd \nu'\,  \nu^{\prime\, 2} \, \frac{\sigma_L(\nu', Q^2)}{\nu^{\prime\, 2} -\nu^2 },
\eeq 
where $\nu_0$ is the threshold for the virtual-photon absorption. The pole contributions in the amplitude (coming from the Born graphs for VVCS) are canceled by the elastic 
contribution to $\sigma_L$. Thus, this dispersion relation has the same form for the non-pole contributions, with $\nu_0$ being the inelastic threshold. 
A non-pole contribution remaining from the Born term contains the Pauli form factor $F_2(Q^2)$, 
\beq 
\eqlab{nonpoleBorn}
T_L (\mbox{non-pole Born}) = - \frac{\pi \al_\mathrm{em} Q^2}{M^3}  F_2^2(Q^2), 
\eeq  
which in the limit of $Q^2\to 0$ gives the anomalous magnetic moment, $F_2(0)=\varkappa$. The rest of  amplitude, in the limit of  $Q^2\to 0$ and $\nu \to 0$, gives
the electric polarizability,
\beq 
\eqlab{LEXTL}
T_L (\mbox{non-Born}) =  4 \pi Q^2 \al_{E1} + O(\nu^2, Q^4)  . 
\eeq
Taking the same low-energy limits for the above dispersion relation gives us the BT sum rule as written in \Eqref{BTsr}. Now, what is wrong with it?

Llanta and Tarrach \cite{Llanta:1978xd} attempted to verify this sum rule in QED by computing the left-hand and right-hand sides to leading order in $\al_\mathrm{em}$, 
and the results differed by a constant. This constant, as they show, is given by the value of $T_L$ at $\nu\to \infty$. This is not surprising if one recalls the 
Sugawara-Kanazava theorem \cite{Sugawara:1961zz}, which basically says that the left-hand side of the dispersion relation [\Eqref{TLDR}] must include the value at
infinity as follows:
\beq 
T_L(\nu, Q^2) - T_L(\infty, Q^2).
\eeq 
This asymptotic contribution is usually assumed to be vanishing, or divergent (in the latter case the dispersion relation requires subtractions). However, in this perturbative calculation 
it apparently is finite. Of course, this test by itself does not invalidate the sum rule, because perturbative QED is invalid at $\nu=\infty$, but it does cast a shadow
over the sum rule applicability. This failure of the sum rule QED (pun intended!) has later been exploited by L'vov \cite{Lvov:1998csf}, who gives more examples and arguments for the sum rule
to be dismissed. 

Our aim here is quite the opposite --- to rehabilitate the BT sum rule. First of all, the value of the amplitude at infinity is there for any dispersion relation and thus may enter any sum rule, 
including the Baldin and other sum rules which are widely used. Empirically there is no way to find out. Theoretically, it is as an artifact, since we usually do not know what happens
at asymptotically high energies. At the same time, it is hard to believe that the use of the sum rule for a low-energy quantity, such as polarizability, depends on physics
beyond the Plank scale. A proper way to show the irrelevance of the value at infinity in QED, or another theory, is to cancel it by a ultraviolet completion set 
at a high-energy scale (where there is no data), and then see how little it contributes to a quantity such as, say, the proton polarizability. 

Instead of doing this program for QED, we shall here identify a perturbative calculation which verifies the BT sum rule exactly.
Incidentally, it is for the proton polarizability.

\section{Validation in baryon $\chi$PT}
Consider the manifestly covariant baryon  $\chi$PT calculation of the proton polarizabilities
to leading order  \cite{Bernard:1991rq,Bernard:1992qa}, and check whether it can be reproduced
by the two sum rules. For the Baldin sum rule, this exercise was done in~\cite{Lvov:1993ex,Pascalutsa:2004wm} by calculating
the tree-level pion photoproduction, see Figs.~\fref{ChargedPion} and \fref{NeutralPion}.
\begin{figure}
    \centering
    \includegraphics[width=0.8\linewidth]{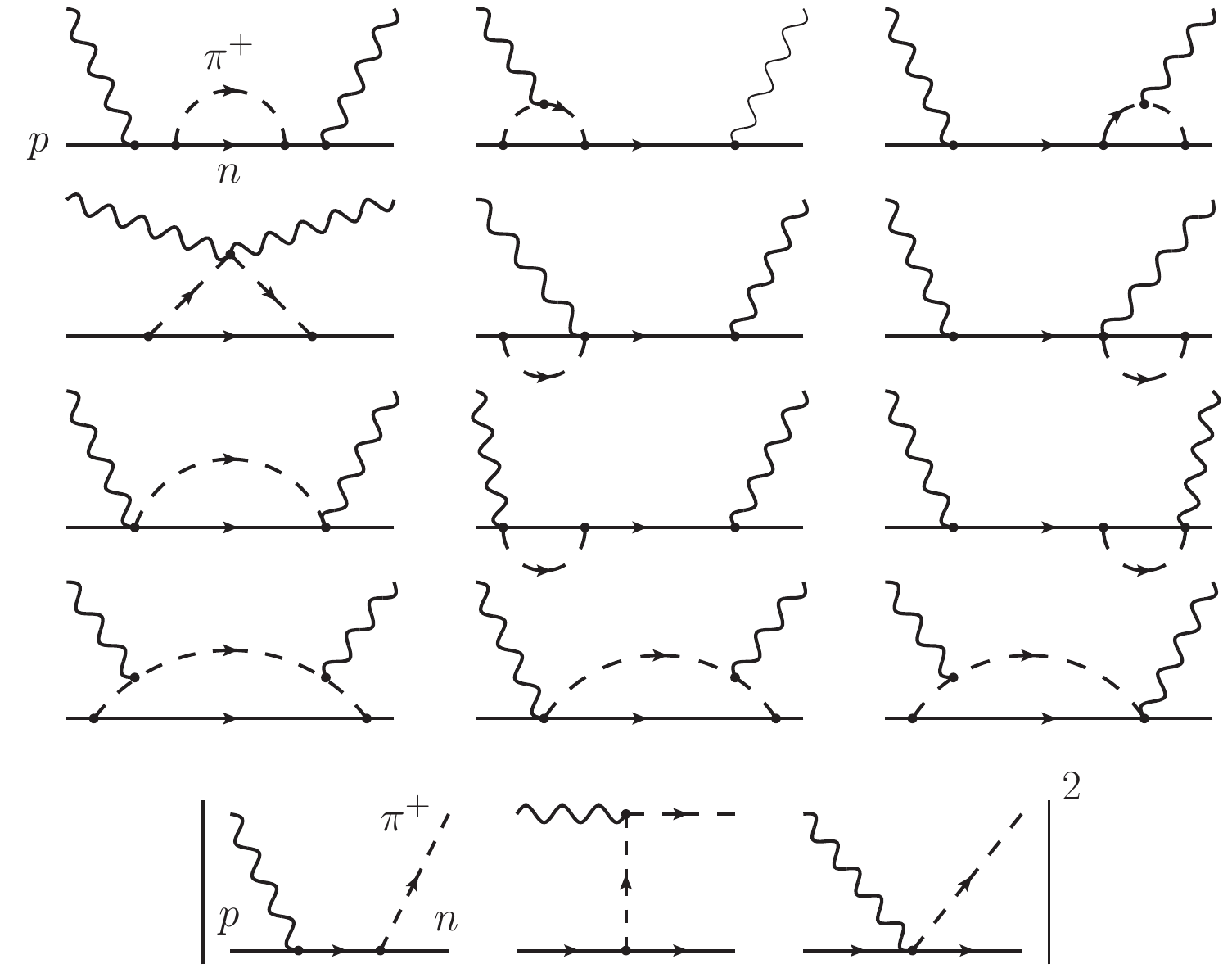}
    \caption{The $\pi^+ n$ contributions to the sum rule, where the loops contribute to the left-hand side and the tree-level cross sections to the right-hand side. }
    \label{fig:ChargedPion}
\end{figure}
\begin{figure}
    \centering
    \includegraphics[width=0.6\linewidth]{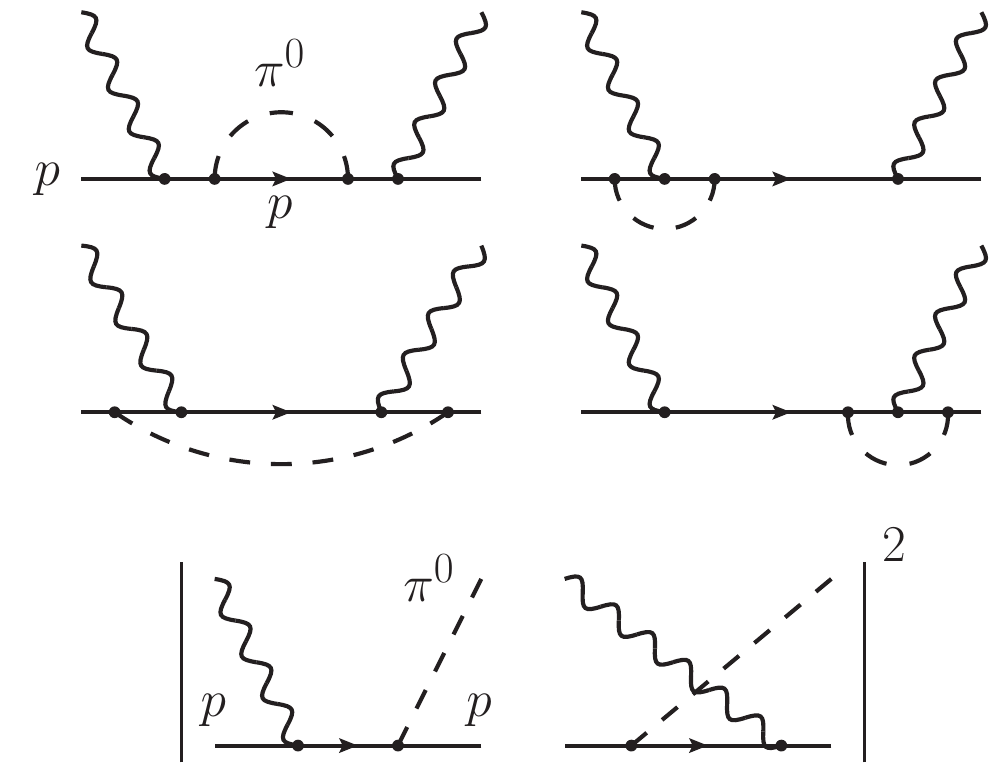}
    \caption{The $\pi^0 p$ contributions to the sum rule, where the loops contribute to the left-hand side and the tree-level cross sections to the right-hand side. }
    \label{fig:NeutralPion}
\end{figure}
Here, we have verified the BT sum rule, by using the longitudinal cross section $\sigma_L$ of charged-pion photoproduction \cite{Lensky:2014dda,Alarcon:2013cba} (note that at this order the 
$\varkappa^2$ contribution to the sum rule is 0). We have also verified explicitly that in this case the VVCS amplitude at infinite energy is vanishing, 
\beq 
T_L^{\pi^+n\text{-loops}}(\infty, Q^2)=0.
\eeq
Hence the sum rule works exactly, and not just up to a constant. 

For the loops involving the neutral pion, corresponding to the $\pi^0 p$-channel on the cross-section side, the amplitude at infinity is   not vanishing. 
As in QED, the constant in the asymptotic value of $T_L$ \footnote{The asymptotic values of the Compton amplitudes were conveniently calculated using Package~X \cite{Patel:2015tea,Patel:2016fam}.},
\beq
T_L^{\pi^0 p\text{-loops}}(\infty,Q^2) = -\frac{\alpha_\mathrm{em}}{12\pi}\frac{g_{\pi N}^2}{M^3}Q^2 + O(Q^4)
\eeq
(here $g_{\pi N} $ is the pion-nucleon coupling constant) comes from the one-particle-irreducible graph, where both photons couple to the Dirac fermion in the loop.
It suggests that this artefact may be handled by a simple ultraviolet completion involving a short-range fermion-fermion interaction. In the proton
case, that role would be played by the leading-order nuclear force.

In any case, the sum rule works, albeit only for the $\pi^+n$ channel it works without the caveat. For an empirical evaluation
of the sum rule, one can safely neglect the value of the amplitude at infinity. Let us see what the sum rule would give 
empirically for the proton.  Unfortunately, we have found only one viable empirical model for $\left[\sigma_L/Q^2\right]_{Q^2=0}$  of the proton --- the MAID \cite{MAID}.
Other parametrizations \cite{Christy:2011,HillerBlin:2019hhz} seem to misbehave in the limit of small $Q^2$; we could not obtain a stable extrapolation to 0.
The MAID, however, provides only one of the contributions to the inclusive cross sections --- the single-pion production ($\pi N$) channel. At least it is the dominant channel at low energies.

We have studied the sum rule integrals as functions of the upper limit of integration, 
\begin{subequations}
\bea
I_\mathrm{BT}(\Lambda) &=& \frac{1}{2\pi^2}\int_{\nu_{0}}^{\Lambda}\dd\nu\left[\frac{\sigma_L(\nu,Q^2)}{Q^2}\right]_{Q^2\to 0},\label{BTint}\\
I_\mathrm{Baldin}(\Lambda) &=& \frac{1}{2\pi^2}\int_{\nu_{0}}^{\Lambda}\dd\nu\frac{\sigma_T(\nu)}{\nu^2}\label{Bint}.
\eea
\end{subequations}
The MAID results are shown in Fig.~\fref{BT_B_integrals} by dashed curves. They can be compared to the solid curves representing the $\chi$PT calculation of the $\pi N$ channel, as explained above.
Note that for the Baldin sum rule the discrepancy between MAID and $\chi$PT is very large because of the $\De$(1232) and other $N^\ast$ resonances. 
For the BT sum rule, the leading-order $\chi$PT describes the empirial MAID cross section rather well, and hence their BT integrals agree at such low cutoffs.

Furthermore, from the $\chi$PT calculation we know that the full  BT integral gives, $I_\mathrm{BT}(\Lambda \to \infty) \simeq 7 \times 10^{-4}$ fm$^3$. Taking into account
the anomalous magnetic moment term, 
\beq 
\frac{\al_\mathrm{em} \varkappa_p^2}{4M_p^3} 
\simeq 0.54\times 10^{-4}\, \mathrm{fm}^3,
\eeq
we obtain the proton electric polarizability of about 7.5 (in the usual units). This can be compared to the PDG value \cite{ParticleDataGroup:2022pth}:
$\alpha_{E1}^p = 11.2\pm0.4$. It is quite plausible that this difference will be diminished by inclusion of other channels, predominately the $\pi \pi N$ channel. One can see  that for the Baldin sum rule the single-$\pi N$-channel value of
about 11.6 is also different from the inclusive result of $14 \pm 0.2$. The relative difference here is smaller than in the BT sum rule,
because apparently the Baldin sum rule converges faster.

\begin{figure}[htb]
    \centering
    \includegraphics[width=0.8\linewidth]{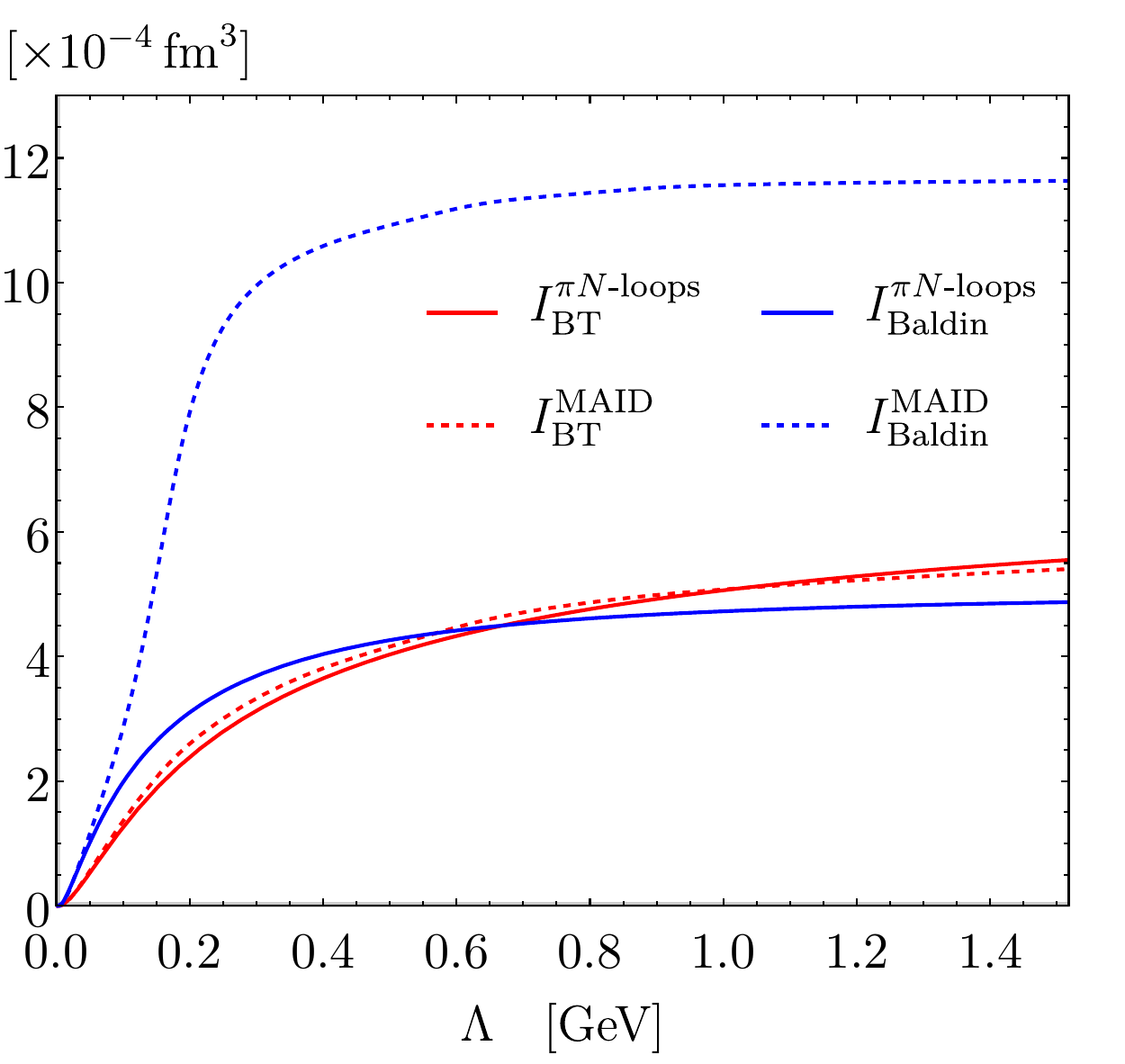}
    \caption{Saturation of the Bernab\'eu-Tarrach and Baldin integrals [Eqs.~\eqref{BTint}, \eqref{Bint}] for the proton, using
    the leading-order $\chi$PT and the empirical MAID model.}
    \label{fig:BT_B_integrals}
\end{figure}

\section{Sum rule for the subtraction function}
\label{sec:subfunSR}
It remains to be seen whether the BT sum rule converges, but if it does, there would be a few
important implications. First of all, it will provide an independent determination of
the nucleon electric polarizability and help to resolve the large
contradiction of two most recent Compton experiments: HI$\gamma$S \cite{Li:2022vnz} versus MAMI \cite{A2CollaborationatMAMI:2021vfy}. Secondly, it will allow to evaluate
the VVCS amplitude $T_L$ via the dispersion relation \eref{TLDR}. This is important for the subtraction-function contribution of the proton polarizability effect in muonic hydrogen. 
The subtraction function sits in the transverse VVCS amplitude $T_1 (\nu, Q^2)$, but can be
calculated via $T_L$ using Siegert's theorem \cite{Siegert:1937yt}, which equates (up to a phase factor) the transverse and longitudinal amplitudes in the limit of $\nu\to iQ$:
\beq
\eqlab{Sieg}
T_1(iQ, Q^2) = - T_L(iQ, Q^2).
\eeq 
Now, this is all that is needed for the subtraction point of $\nu = i Q$ \cite{Hagelstein:2020awq}. More usual is the subtraction at $\nu=0$, which implies
the following dispersion relation for $T_1$:
\beq 
\eqlab{T1DR}
T_1(\nu, Q^2 )= T_1(0, Q^2) + \frac{2\nu^2}{\pi} \int_{\nu_0}^\infty \dd \nu' \, \frac{\sigma_T(\nu', Q^2)}{\nu^{\prime\, 2} -\nu^2 } .
\eeq 
Combining it with the dispersion relation for $T_L(iQ,Q^2)$ and the Siegert theorem, the conventional subtraction function has the following expression:
\beq 
\eqlab{subSR}
T_1(0, Q^2 ) = \frac{2}{\pi} Q^2 \int_{\nu_0}^\infty \frac{\dd \nu}{\nu^2 + Q^2 } \left[ \sigma_T - \frac{\nu^2}{Q^2} \sigma_L \right](\nu, Q^2).
\eeq 
We have verified this sum rule exactly in the $\chi$PT example above, including the charged-pion channel. Note that at this order we only verify the polarizability contribution
and not any of the possible non-pole VVCS contributions coming from the Born term (expressed by the elastic form factors).

\begin{figure*}[htb]
    \centering
    \begin{tabular}{cc}
    \includegraphics[width=0.45\linewidth]{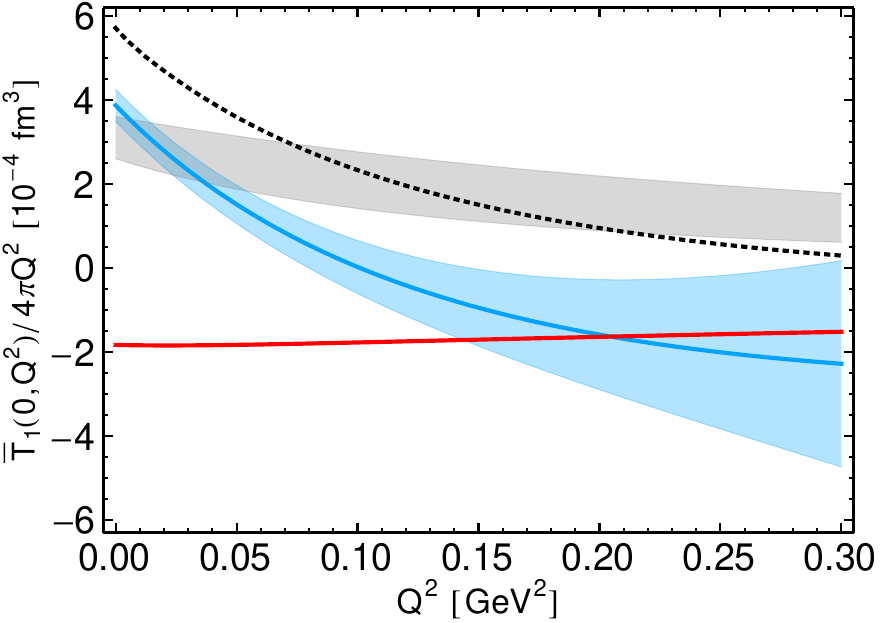}
    &
    \includegraphics[width=0.45\linewidth]{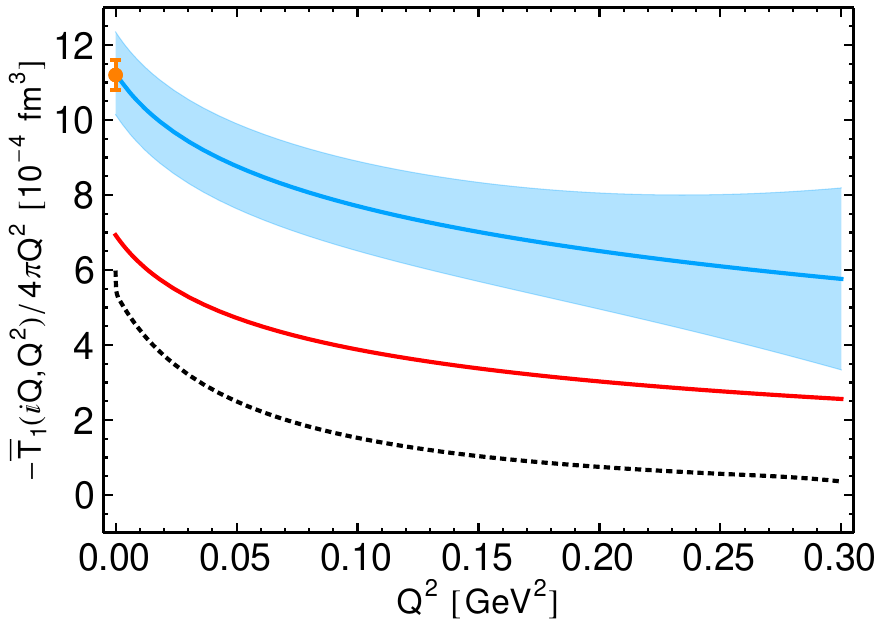}
    \end{tabular}
    \caption{Non-Born part of the subtraction functions, $\ol T_1(0,Q^2)$
    (left panel) and $\ol T_L(iQ,Q^2)$ (right panel), evaluated with MAID \cite{MAID} through Eqs.~\eref{subSR} and \eref{TLDR} (black dotted). For comparison we show: the heavy-baryon $\chi$PT calculation~\cite{Birse:2012eb} (gray band); 
the next-to-leading-order $\chi$PT calculation~\cite{Lensky:2014dda,Alarcon:2020wjg} (blue solid with a band); the leading $\chi$PT $\pi N$-loop contribution (red solid);
At the real photon point, the PDG value for $\alpha_{E1}=(11.2\pm 0.4)\times 10^{-4}~\text{fm}^3$ \cite{ParticleDataGroup:2022pth} is shown. }
    \label{fig:subtraction}
\end{figure*}

In \Figref{subtraction}, we show the non-Born part of the VVCS amplitudes as functions of $Q^2$ evaluated through the integrals on the right-hand side of \Eqref{subSR} (left panel) and \Eqref{TLDR} (right panel)  using MAID \cite{MAID}. 
In order to obtain the non-Born part, from the above dispersion relations, one has to subtract the non-pole Born parts. For $T_L$, it is given by \Eqref{nonpoleBorn}. In the case of $T_1(0,Q^2)$ evaluated through \Eqref{subSR}, one has to add \Eqref{nonpoleBorn}. At first glance, this might look counterintuitive, since the non-pole Born part of $T_1(\nu,Q^2)$ is given by $T_1(\text{non-pole Born})=-4\pi \al_\mathrm{em}\,F_1^2(Q^2)/M$. The difference comes from the mismatch of the non-pole parts in \Eqref{Sieg}: this equality is valid for the full Born amplitudes, but not separately for the pole and non-pole contributions. As the result, we have the following expressions for the non-Born parts of the subtraction functions, plotted in \Figref{subtraction}:
\begin{subequations}
\bea
\overline T_1(0, Q^2 ) &=&-\frac{\pi \al_\mathrm{em} Q^2}{M^3}  F_2^2(Q^2)\\
&&+\frac{2}{\pi} Q^2 \int_{\nu_0}^\infty \frac{\dd \nu}{\nu^2 + Q^2 } \left[ \sigma_T - \frac{\nu^2}{Q^2} \sigma_L \right](\nu, Q^2),\nn\\
\overline T_L (iQ, Q^2) &=&\frac{\pi \al_\mathrm{em} Q^2}{M^3}  F_2^2(Q^2)\\
&&+ \frac{2}{\pi} \int_{\nu_0}^\infty \dd \nu\,  \nu^{ 2} \, \frac{\sigma_L(\nu, Q^2)}{\nu^2 +Q^2 } = - \overline T_1 (iQ, Q^2),\nn
\eea 
\end{subequations}
where $\nu_0$ is again the inelastic threshold. In calculating the Pauli form factor ($F_2$) contribution,  we are using the empirical parametrizations of the nucleon form factors from Ref.~\cite{Bradford:2006yz}. 

In the limit of $Q^2=0$, the $\ol T_L(iQ,Q^2)/4\pi Q^2$ amplitude yields $\alpha_{E1}$, cf.\ \Eqref{LEXTL}, whereas $\ol T_1(0,Q^2)/4\pi Q^2$ yields $\beta_{M1}$. Their sum is consistent with the MAID evaluation of the Baldin sum rule shown in \Figref{BT_B_integrals}.

Besides the data-driven evaluations of the subtractions functions based on the MAID pion-production cross sections, we show the leading and next-to-leading $\chi$PT predictions of the amplitudes, and the PDG values of the proton polarizabilities.

The figure (right panel) shows that the agreement of B$\chi$PT prediction with the empirical value of $\alpha_{E1}^p$ is only achieved at the next-to-leading order, where the $\pi \Delta$-loops are included. 
This would correspond to the inclusion of $\pi \De$-production channel in $\si_L$, which goes beyond 
MAID. Hence one would need to include at least the 
two-pion production channel to saturate the BT sum rule
in a data-driven evaluation.

\section{Validation in the parton model}

Another interesting regime where the relations \eref{TLDR} and \eref{subSR} could be checked theoretically is the domain of Bjorken scaling. In this limit the virtuality $Q^2$ and the energy $\nu$ of the incoming photon are taken to be very large, while preserving the finiteness of the Bjorken variable $x=Q^2/2M\nu$. The Bjorken scaling implies the validity of the perturbative expansion of QCD, with the leading-order contribution given by the na\"ive parton model.

In this model the deep inelastic scattering off the proton is described by the scattering off the individual partons (quarks) with electric charges $e_q$ and parton distribution functions $f_q(x)$. The structure functions $F_1$ and $F_2$ are then given by (see, e.g., Sec.~18.5 in \cite{Peskin:1995ev})
\begin{subequations}
\bea
F_1(x,Q^2) && = \frac{M}{\al_\mathrm{em}}\sum_q e_q^2 f_q(x),\\
F_2(x,Q^2) && = 2x F_1(x,Q^2),\eqlab{CallanGross}
\eea
\end{subequations}
while the corresponding forward Compton amplitudes are: 
\begin{subequations}
\bea
T_1(x,Q^2) &=& 2\int_0^1\frac{d\xi}{\xi} \frac{\sum_q e_q^2 f_q(\xi)}{(\frac{x}{\xi})^2-1-i\epsilon},\\
T_2(x,Q^2) &=& 8\frac{M^2x^2}{Q^2}\int_0^1\frac{d\xi}{\xi}\frac{\sum_q e_q^2 f_q(\xi)}{(\frac{x}{\xi})^2-1-i\epsilon}.
\eea
\end{subequations}
The longitudinal structure function $F_L$ is given in this model by
\bea
F_L(x,Q^2) &=& \left(1+\frac{4M^2x^2}{Q^2}\right)F_2(x,Q^2)-2xF_1(x,Q^2)\nn\\
&=& \frac{4M^2x^2}{Q^2}F_2(x,Q^2)\\ 
&=& \frac{M^3}{\al_\mathrm{em}}\sum_q e_q^2 f_q(x)\frac{8x^3}{Q^2},\nn
\eea
and the corresponding Compton amplitude is
\bea
T_L(x,Q^2) &=& \left(1+\frac{\nu^2}{Q^2}\right)T_2(x,Q^2) -T_1(x,Q^2) \nn\\
&=& T_2(x,Q^2)
\eea
This relation immediately implies the convergence of the unsubtracted dispersion relation for $T_L$, 
as long as the unsubtracted relation is valid for $T_2$.

A remarkable feature of the parton model is the Callan-Gross relation \cite{Callan:1969uq} in \Eqref{CallanGross}
or, equivalently, in terms of the cross sections,
\beq
\sigma_L = \frac{Q^2}{\nu^2}\sigma_T,
\eeq
which shows that $\sigma_L$ falls with energy considerably faster than $\sigma_T$.
Writing the unsubtracted dispersion relations for the Compton amplitudes in the parton model,
\begin{subequations}
\bea
T_1(x,Q^2) &=& \frac{2\al_\mathrm{em}}{M}\int_0^1 \frac{d\zeta}{\zeta^3}\frac{x^2 F_1(\zeta,Q^2)}{(\frac{x}{\zeta})^2-1-i\epsilon},\\
T_2(x,Q^2) &=& \frac{4M\al_\mathrm{em}}{Q^2}\int_0^1 \frac{d\zeta}{\zeta^2}\frac{x^2 F_2(\zeta,Q^2)}{(\frac{x}{\zeta})^2-1-i\epsilon},
\eea
\end{subequations}
one can deduce that the amplitude $T_2$ (and, consequently, $T_L$) satisfies the unsubtracted dispersion relation, but $T_1$ does not. The latter, however, satisfies the once-subtracted dispersion relation with vanishing subtraction function. Using the Callan-Gross relation one trivially verifies the sum rule for the subtraction function $T_1(0,Q^2)$, given by \eref{subSR}. Hence, we conclude that the dispersion relation for $T_L$ \eref{TLDR} as well as the sum rule for the subtraction function \eref{subSR} hold exactly in the na\"ive parton model.

\section{Conclusion}
The BT sum rule for the nucleon electric polarizability is the same level of validity as the commonly used Baldin sum rule, despite an appreciably worse
convergence. We have established at least one simple example where the BT sum rule passes the perturbative test -- the $\pi^+ n$ channel at leading-order $\chi$PT.
In other cases, most notably the leading-order QED \cite{Llanta:1978xd}, the sum rule holds up to a constant yielded by an unphysical behavior of the VVCS amplitude at infinite energy. 

The na\"ive parton model also verifies the unsubtracted dispersion relation for the longitudinal amplitude, 
crucial for the validity of the BT sum rule.

As an implication of the good BT sum rule, we have derived a sum rule for the subtraction function \eref{subSR}, which will allow for a fully data-driven evaluation of the
proton polarizability contribution to the Lamb shift of (muonic-)hydrogen. To carry out such an evaluation, we need high-quality and precision parametrization of proton $\sigma_L(\nu,Q^2)$  [equivalently, the longitudinal structure function $ F_L(x,Q^2)$] High-quality parametrizations of $\sigma_L$ are needed as well to determine the proton electric polarizability
from the BT sum rule itself. We expect that the inclusion of the two-pion production channel, in addition to the single-pion production parametrized by MAID, will be sufficient to saturate the sum rule to a large extent.

\acknowledgments
We thank Marc Vanderhaeghen for stimulating conversations and  insights into current parametrizations of the longitudinal structure function. We are grateful to Michael Birse and Martin Hoferichter for useful remarks on the manuscript.

This work is supported by the Deutsche Forschungsgemeinschaft (DFG) within the Research Unit FOR 5327 “Photon-photon interactions in the Standard Model and beyond - exploiting the discovery potential from MESA to the LHC” (grant 458854507) and
through the Emmy Noether Programme (grant 449369623). 


%

\end{document}